\documentclass{article}
\usepackage{epsfig}
\usepackage{graphicx}
\usepackage{fancyhdr}

\catcode`\^^M=10 

\title{Efficient Binary and Run Length Morphology and its Application to Document Image Processing}
\author{Thomas M. Breuel \\ DFKI and U. Kaiserslautern \\ tmb@iupr.dfki.de}

\begin{document}
\date{}
\maketitle

\begin{abstract}
This paper describes the implementation and evaluation of an open source
library for mathematical morphology based on packed binary and
run-length compressed images for
document imaging applications.  Abstractions and patterns useful
in the implementation of the interval operations are described.
A number of benchmarks and comparisons
to bit-blit based implementations on standard document images are provided.
\end{abstract}

\section{Introduction}

Binary morphology is an important and widely used method in document image analysis,
useful for tasks like 
image cleaning and noise removal, \cite{Ye01cleaning}
layout analysis, \cite{Wong82}
skew correction,  \cite{najman04skew}
and
text line finding. \cite{das01skew}

Real-world document analysis systems currently primarily rely on bit blit-based
implementations.

Practical implementations take advantage of separability and
logarithmic decomposition of rectangular structuring elements
\cite{rvdb92bitblit,bloomberg02bits,najman04skew}.

\par

This technical report describes a binary morphology library containing both
a run-length and a packed binary implementation of morphological operations.

A number of the methods described in this paper are very similar to
methods described in the literature \cite{piper89rle,rvdb92bitblit},
although the library was developed indepently of that literature.

The paper will not provide a detailed discussion of the similarities
and differences of the algorithms described in this memo to those in
the literature\footnote{Comments and additional references
to prior work would be appreciated, however.}

This memo does provide a number of benchmarks that should help
practitioners choose good algorithms for their particular applications.

\par

We note that, in addition to run length and packed binary methods, a
number of other methods have been described in the literature.

Binary mathematical morphology with convex structuring elements can
be computed by propagation of distances on the pixel grid using
a dynamic programming algorithm \cite{vincent92mmalg} (brushfire algorithms).

Another class of algorithms is based on contours \cite{verwer88contour} and
loop and chain methods \cite{vincent92mmalg}.

The van Herk/Gil-Werman algorithms
\cite{vanherk92filters,gil93filters,gil02efficient} have constant
per-pixel size overhead for grayscale morphology, and binary morphology
can be viewed as a special case.

Another class of algorithms is taking advantage of {\em anchors},
\cite{mvd05anchors}.

Although some of these algorithms are competitive for gray scale morphology,
they have not been demonstrated to be competitive with high quality
bit blit-based implementations for binary morphology on packed binary
representations \cite{bloomberg02bits}.

\par

Some authors have looked again at grayscale morphology, using more complex
intermediate representations \cite{mvd02rectangular}.

It remains to be seen how such algorithms compare to the algorithms
in this paper, both in performance and storage; we will not be addressing
that question here.

\par

Bit blit-based implementations at their lowest level take advantage of
operations that are highly efficient on current hardware because they are
used as part of many different algorithms and display operations: their
running time grows quadratically in the resolution of the input image;
they do not take advantage of coherence in the input image--an almost
blank image takes the same amount of time to process as a highly detailed
image; and operations that need to take into account the coordinates
of individual pixels (e.g., connected component labeling) often need to
decompress (at least on the fly) or use costly pixel access functions.

\par

We will mostly limit ourselves in this paper to the development of
morphological operations involving rectangular structuring elements.

These are by far the most common operations in document image
analysis. 

However, the run-length method can also be used for implementing
morphological operations for arbitrary masks; algorithms and
performance will be given in a separate paper.

\par

Converting between run length and non-run length representations can
be carried out fairly quickly, so we also have the option of mixing
run-length and bitmap representations.

However, many binary image processing algorithms can be implemented
directly on run length images. 

In fact, prior work in image processing
on the line adjacency graph and algorithms operating on it are
directly transferable.

We therefore briefly discuss a number of these algorithms.

Taken together with the binary morphology operations in this paper,
they allow complete binary image processing pipelines to be built
on run length images, with no conversion costs.

\par

Finally, we give benchmarks and comparisons with the Leptonica
library, an open source library for morphological image processing.
It has comparatively good performance, uses well-documented
algorithms, and is used in several large-scale document analysis systems.

\section{Run Length Image Coding}

Run-length image representations have a long history in image
processing and analysis.

They have been used, for example, for efficient storage of binary and
color images and for skeletonization of large images.

\par

Consider a 1D boolean array $a$ containing pixel values $0$ and $1$ at
each location $a_i$.

The run length representation $r$ is an array of intervals
$r_1=[s_1,e_1], \ldots, r_n=[s_n,e_n]$ such that $a_i=1$ iff $i\in
r_j$ for some $j$ and $e_i < s_{i+1}$.

\par

The 2D run-length representation we are using in this paper is a straight-forward,
extension to 2D that treats the two coordinates asymmetricaly;

in particular, the binary image\footnote{
This paper and our library uses PostScript/mathematical conventions, with
$a_{0,0}$ representing the bottom left pixel of the image.
}
$a_{ij}$ is represented as a sequence of
one-dimensional run-length representations $r_{ij}$, such that for any fixed
$i_0$, the 1D array $a_j = a_{i_0,j}$ is represented by the 1D runlength
representation $r_j = r_{i_0j}$.

\par

Even algorithms that are not explicitly using run length
representations are often still implicitly manipulating runs of pixels
internally; for example, the usual connected component labeling
algorithm internally considers neighborhood relations between runs of
pixels.

An extended version of 2D run-length representations has been used as
the line adjacency graph (LAG); it adds a graph structure encoding
neighborhood relations between runs to the basic run length encoding;
in our algorithms, these neighborhood relations are simply inferred
dynamically.

\par

In practice, since all the algorithms described in this paper access
runs sequentially, both linked lists or extensible arrays with
exponential doubling can be used to represent the runs of each line;

our implementation uses extensible arrays with exponential doubling,
which results in fewer calls to the memory allocator, less average
memory usage, and better locality of reference than linked list
representations.

In our current implementation, each run is represented as a pair of 16
bit integers, allowing images up to approximately $65535 \times 65535$
to be represented; other, more efficient coding schemes are possible
(e.g., using a Unicode-like variable length encoding).

Note that, in the worst case, that of alternating black and white
pixels, the run-length representation may be up to 16 times bigger
than a packed bit representation, or a factor of two compared to a
one-byte-per-pixel representation.

\par

On the other hand, run-length encoded images scale linearly with
image resolution, rather than quadratically.

That is, a 1200 dpi binary image takes approximately 4 times as much
space than a 300 dpi binary image using run-length encoding, while a
packed bit image would take 16 times as much space.

\section{Morphological Operations}

Because of the asymmetry in the two dimensions of the 2D run-length
representation we are using, morphological operations behave
differently in the $x$ and $y$ direction in run-length
representations.

An analogous asymmetry is found in bit-blit operations, in which the
bits making up image lines are packed into words, and a list of lines
represents the entire image.

There are multiple possible approaches for dealing with this issue.

First, we can implement separate operations for horizontal and
vertical operations.

Second, we can implement only the within-line operations and
then transform the between-line operations into within-line operations
through transposition.

For separable operations, the second approach is often the easier one.

Therefore, an erosion with a rectangular structuring element of size
$u \times v$ can be written as:\footnote{
Our convention is output arguments before input arguments, and the various
procedures modify the image in place.
}
\begin{verbatim}
function erode2d(image,u,v)
  erode1d(image,u)
  transpose(image)
  erode1d(image,v)
  transpose(image)
end
\end{verbatim}

\section{Within-Line Morphological Operations}

There are four basic morphological operations we consider:
erosion, dilation, opening, and closing.

One-dimensional opening and closing are the easiest to understand.

Essentially, a one-dimensional opening with size $u$ simply deletes
all runs of pixels that are less than size $u$ large, and leaves all
others untouched:\footnote{We are using 1-based arrays in the pseudo-code.}

\begin{verbatim}
void open1d(image,u) {
  for i in 1,length(image.lines) do
    line = image.lines[i]
    filtered = []
    for j in 1,length(line.runs) do
      if runs[j].width() >= u then
        filtered.append(line.runs[j])
      end
    image.lines[i] = filtered
  end
end
\end{verbatim}

A one-dimensional closing with size $u$ deletes all gaps that are smaller
than size $u$, joining the neighboring intervals together.

It can either be implemented directly, or it can be implemented in terms of
complementation and opening\footnote{
To simplify boundary conditions, we are using the notation
{\tt exp1 or exp2} to mean use the value of {\tt exp1} if it is defined, 
otherwise use {\tt exp2}.
}
\begin{verbatim}
function complement(image)
  for i in 1,length(image.lines) do
    line = image.lines[i]
    filtered = []
    last = 0
    for j in 1,length(line.runs) do
        run = line.runs[j]
        newrun = make_run(last,run.start)
        filtered.append(newrun)
        last = run.end
    end
    filtered.append(make_run(last,maxint))
    image.lines[i] = filtered
  end
end

function close1d(image,u)
  complement(image)
  open1d(image)
  complement(image)
end
\end{verbatim}

Note that openings and closing are not separable, so we cannot use
these implementations directly for implementing true 2D openings and closings;
for that, we have to combine erosions and dilations.

However, even as they are, these simple operations are already useful
and illustrate the basic idea behind run-length morphology: run-length
morphology is selective deletion and/or modification of pixel runs.

\par

The most important operation in run-length morphology is one-dimensional erosion.

Like one-dimensional opening, we walk through the list of runs, but
instead of only deleting runs smaller than $u$, we also shrink runs
larger than $u$ by $u/2$ on each side (strictly speaking, for erosions
on integer grids, we shrink by $\hbox{floor}(u/2)$ on the left side
and $u - \hbox{floor}(u/2)$ on the right side during erosions), and
use the opposite convention for dilations).

In pseudo-code, we can write this as follows:

\begin{verbatim}
function erode1d(image,u)
  for i in 1,length(image.lines) do
    line = image.lines[i]
    filtered = []
    for j in 1,length(line.runs) do
      if runs[j].width() >= u then
        start = runs[j].start+u/2
        end = runs[j].end-u/2
        filtered.append(make_run(start,end))
      end
    image.lines[i] = filtered
  end
end
\end{verbatim}

As with opening/closing, dilation can be implemented directly or via complementation:

\begin{verbatim}
function dilate1d(image,u)
  complement(image)
  erode1d(image)
  complement(image)
end
\end{verbatim}

In terms of computational efficiency, obviously, all these operations
are linear in the total number of runs in the image.

\section{Efficient Transpose}

Transposition means that we need to construct runs of pixels in the
direction perpendicular to the current run-length encoding.

A simple way of transposing is to essentially decompress each
run individually and then accumulate the decompressed bits
in a second run length encoded binary 
image \cite{anderson88transpose,misra99transpose}.

For this, we maintain an array of currently open runs in each line of
the output image and iterate through the runs of the current line in
the input image.

For the range of pixels between the runs of the current line in the input image,
we finish off the corresponding open runs in the output image.

For the range of pixels overlapping the runs of the current line in
the input image, we start new runs for lines where runs are not
currently open and continue existing open runs for lines where runs
are currently open.

In terms of pseudo code, that looks as follows:

\begin{verbatim}
function transpose_simple(image) 
  output = make_rle_image()
  open_runs = make_array(new_image_size)
  for i = 1,length(image.lines) do
    line = image.lines[i]
    last = 1
    for j=1,length(line.runs) do
      run = line.runs[j]
      for k=last,run.start do
        newrun = make_run(open_runs[k],i)
        output.lines[k].append(newrun)
        open_runs[k] = nil
      end
      for k=run.start,run.end do
        if open_runs[k] == nil then
            open_runs[k] = i
        end
      end
      last = run.end
    end
  end
  ... finish off the remaining runs here ...
end
\end{verbatim}

This simple algorithm is usable, but it does not take advantage of
the coherence between lines in the input image.

To take advantage of that, we need a more complicated algorithm; the
algorithm is somewhat similar to the rectangular sweeping algorithm
used for finding maximal empty rectangles \cite{bjf90}.

\par

The basic idea behind the transposition algorithm is to replace the
array of open runs in the above algorithm with a list of runs, each of
which represents an open run in the perpendicular direction.

\begin{figure}
\includegraphics[width=5in]{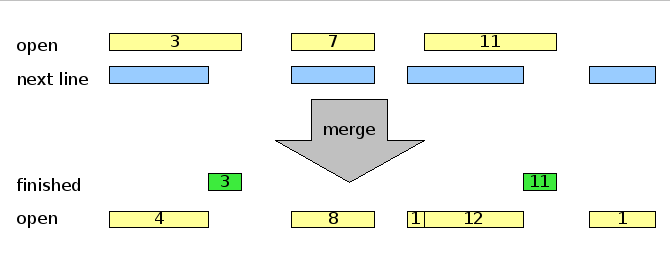}
\caption{\label{fig-runs}
The figure illustrates a merge step during the transposition step.
The algorithm maintains a list of open intervals and information about
how many steps that interval has been open for.  It then considers the next
run-length encoded line in the input.  Ranges in the input that do not overlap
any intervals in the new line are finished and give rise to runs in the output.
Ranges in the input that overlap runs in the next line give rise to intervals
in the open line that have their step number incremented by one.  Ranges in
next line that do not correspond to any range in the list of open intervals
give rise to new intervals with their step values initialized to one.
}
\end{figure}

This is illustrated in Figure~\ref{fig-runs}.

The actual inner loop is similar to the algorithm shown above for the
per-pixel updating, but because of the 13 possible relationships
between two closed intervals, the inner loop contains
a larger case statement; this will not be reproduced here.

This new run length transposition algorithm speeds up the overall
operation of the binary morphology code several-fold relative to the
simple decode-recode implementation.

\section{Between Line Boolean Operations}

A bit blit based implementation of mathematical morphology uses as its
primitive an operation that performs a logical operation (AND, OR, XOR,
AND-NOT, OR-NOT, NAND, NOR, etc.) between the pixels of two images,
allowing for relative shifts (see also \cite{young81binary}).

We can implement the same operations in our library.

The general idea is to consider the runs in the two source lines from
left to right and merge them together.

For example, if the operation is AND, then this means deleting any run
in either of the two input lines that does not overlap a corresponding
run in the other line.

Because of the 13 possible relationships between intervals, this is
fairly complicated.

However, there is a pair of useful abstractions that greatly simplifies the
implementation of these kinds of interval merging operations.\footnote{
A similar technique can be applied to the transpose operation above,
but that operation was written by considering the different possible
cases directly.  }

Basically, instead of considering lines as collections of runs, we
consider them as collections of transitions from black to white and from
white to black.

We operate on those collections using two simple abstractions:

\begin{itemize}

\item A {\tt TransitionSource} returns locations of transitions in ascending order, 
    indicating for each transition whether it is from black to white or from white to black.

\item A {\tt TransitionSink} accepts locations of transitions in ascending order
and re-assembles them into intervals.

\end{itemize}

With these abstractions, the main loop of {\tt line\_and} 
becomes simply (in C++):

\begin{verbatim}
TransitionSink sink(out,total);
TransitionSource src1(l1,0);
TransitionSource src2(l2,offset2);
int where = MININT;
bool b1 = false;
bool b2 = false;
while(src1 || src2) {
    if(src1.coord()<src2.coord()) {
        b1 = src1.value();
        where = src1.coord();
        src1.next();
    } else {
        b2 = src2.value();
        where = src2.coord();
        src2.next();
    }
    sink.append(where,b1&&b2);
}
\end{verbatim}

The simplification of the code results from the fact that the main loop
becomes a simple ordered merge of lists of numbers, and the sink data
structure takes care of the different cases; for example if it receives
a sequence of transitions $(x_0,F),(x_1,T),(x_2,T),(x_3,F),(x_4,F)$,
it will generate a run from $x_1$ to $x_3$.

Boundary conditions are additionally simplified because the {\tt TransitionSource}
returns a big integer after running out of locations, which eliminates separate
code after the main loop to finish off the unfinished line when the other line
has run out of transitions.

\section{Efficient Binary Decompositions}

Binary decomposition of linear structuring elements is a widely used technique
for real-world, fast binary morphology.

For reviews of the technique, see \cite{rvdb92bitblit,bloomberg02bits,najman04skew}.

We can combine the ``bit blit-like'' operations above with binary decomposition for an alternative approach to binary morphology on run-length images.

\par

Generally, a linear structuring element of length $n$ can be decomposed
into

\begin{equation}
nS = S \oplus E(S) \oplus E(2S) \oplus \cdots \oplus E(2^{m-1}S) \oplus E((n-2^{m-1})S)
\end{equation}

A simple example of this is the decomposition of the $8L_0$ structuring element\cite{najman04skew}:

\begin{equation}
8L_0 = \{\bullet\bullet\bullet\} 
    \oplus \{\bullet\circ\bullet\} 
    \oplus \{\bullet\circ\circ\circ\bullet\} 
    \oplus \{\bullet\circ\circ\circ\circ\circ\circ\circ\bullet\}
\end{equation}

This involves 9 pixels, and hence 9 blit operations in a straightforward bit blit
implementation. 

This represents a significant savings over the original 19 pixels of the linear structuring
element, but it does not represent the optimal decomposition of a structuring element 
of 19 pixels, which requires just 5 blit operations:

\begin{verbatim}
width = 1;
while(2*width<r) {
    bits_and(image,image,width,0);
    width *= 2;
}
if(width<r) bits_and(image,image,r-width-1,0);
\end{verbatim}

To center the operation properly, we need to shift the image prior to these operations.

The overall cost of decomposing a line structuring element therefore is a shift
operation plus $\lceil \log r \rceil$, where $r$ is the width of the structuring
element in pixels.

In the run length morphology library, this approach is applied to the between-line
morphological operations.

\par

An alternative loop avoids the initial shift, which causes some
undesirable behavior at the image border.  The idea is to perform
exponential doubling to cover at least half of the right half of the
structuring element, and then finish of the operation with three more
operations (overall, this requires one extra blit relative to the simple
version above):

\begin{verbatim}
width = 1;
while(2*width<r/2) {
    bits_and(image,image,width,0);
    width *= 2;
}
if(width<r/2) bits_and(image,image,r/2-width-1,0);
bits_and(image,image,-(r-r/2),0);
if(width<r-r/2) bits_and(image,image,-(r-r/2)+width,0);
\end{verbatim}.

\section{2D Morphological Operators for Rectangular Masks}

Given the three operations developed above, within-line morphological
operations, transpose, between-line binary operations, and logarithmic
binary decomposition, as well as the separability property of rectangular
masks, we now have various different ways of composition 2D morphological
operators using rectangular structuring elements:

\begin{itemize}

\item {\it ``brute force'' implementation}: this simply performs a shifted AND
    or OR operation for each pixel in the mask, quite analogous to a brute-force
    bit blit implementation; this is a slow operation useful for reference
    and verification

\item {\it transpose and within-line operations}: we first perform morphology
    along the within-line direction, then transpose, then perform morphology
    again, and then transpose back

\item {\it within-line and between-line operations}: we perform
    morphology along one direction using the within-line operators, and
    along the other direction using logarithmic decomposition and
    between-line operations; this can be carried out in either order

\end{itemize}

It is not {\it a priori} obvious which of these choices is the most
efficient, but it turns out experimentally that the last one works fastest
for document images.

Furthermore, it is important to carry out the within-line operations before
the between-line operations because the former are far more efficient when
coping with images with many runs.

We will return to this issue in the experimental section.

\section{Morphology with Arbitrary Masks}

Many structuring elements in practice are handled as arbitrary bitmasks, using
a straightforward loop such as:

\begin{verbatim}
for(i=0;i<w;i++) for(j=0;j<h;j++) {
    if(element(i,j)==0) continue;
    bits_and(result,image,-i+cx,-j+cy);
    count++;
}
\end{verbatim}

Obviously, loops like that grow linearly in the number of pixels in the structuring element.

\par

Fortunately, for run-length encodings, we can perform these operations a run
at a time, rather than a pixel at a time.

Essentially, for each run in the mask, we perform the operation on a copy of
the image, then apply the result to an accumulator image, which we finally
return.

\begin{verbatim}
for(run in horizontal_runs(element)) {
    temp = copy(image);
    erode_line_horizontal(temp,run.x,run.y0,run.y1);
    bits_and(result,temp)
    count++;
}
\end{verbatim}

Of course, for run length morphology, we can carry out \verb|erode_line_horizontal|
directly.

\par

For bit blit morphology, the naive implementation would use binary decomposition
to compute \verb|erode_line_horizontal| separately for each run.

For completeness (and because it is implemented in the companion bit blit-based
operations) let us observe that
such an approach involves many unnecessary copies and recomputations
of intermediate results; the overall complexity is 
$\#runs \cdot \log(\hbox{max\_run\_width})$.

A better way to perform decomposition of arbitrary masks is to accumulate longer
and longer horizontal structuring elements and apply them as needed.

The following simple pseudocode illustrates the idea (it assumes
that the structuring element has been flipped prior to the computation; 
\verb|maxwidth| is the maximum run width):

\begin{verbatim}
temp = copy(image);
for(width=1;width<maxwidth;width*=2) {
    for(run in runs_of_element) {
        run_width = run.y1-run.y0;
        if(run_width >= width && run_width < 2*width) {
            bits_and(image,temp,run.x,run.y0);
            if(run_width>width) bits_and(image,temp,run.x,run.y1-run_width-1);
        }
    }
}
\end{verbatim}

For a circle of radius $r$, this involves at most $2r + \lfloor\log\hbox{maxwidth}\rfloor$ 
calls to the \verb|bits_and| function, plus a copy.

A separate shift is not needed to center the result since the runs themselves can be
offset appropriately.

\section{Scaling, Skewing, and Rotation}

Scaling, skewing, and rotation are other important operations in document image
analysis, used during display and skew correction.

\par

Scaling can be implemented by scaling the coordinates of each
run and scaling up or down the array holding the lines by deleting
or duplicating line arrays.

Scaling can also be implemented as part of the conversion into an
unpacked representation (as required by, for example, window systems).

\par

Skew operations can be implemented within each line by shifting the
start and end values associated with each run.  

Bitmap rotation by arbitrary angles can then be implemented by the
usual decomposition of rotations into a sequence of horizontal and 
vertical skew operations, using
successive application of transposition, line skewing, and
transposition in order to achieve skews perpendicular to the lines in
the run-length representation.

We note that this method differs substantially from previously published
rotation algorithms for run length encoded images \cite{zhu95rotation,au02rotation}.

\section{Morphology with Lines at Arbitrary Orientations}

Finally, let us look at one more case related to lines: morphological operations with
linear structuring elements at arbitrary angles.

A number of papers have been written about this, using different approaches.
Oddly enough, few if any of the papers appear to reference the most obvious approach:
bitmap rotation followed by an axis-aligned structuring element; this would at least
be an important control experiment \cite{soille96lines,soille01lines}:

\begin{verbatim}
bits_erode_line(image,r,angle) {
    bits_rotate(image,angle);
    erode_line_horizontal(image,2*r);
    bits_rotate(image,-angle);
}
\end{verbatim}

The rotations can be composed from three skew operations, as for general bitmap
rotations.

In fact, the same approach also works for rotated rectangles.

\par

However, a rotation of a horizontal line segment will give rise to a rotated line
segment whose bits differ slightly from the bits generated by rendering a digital
line segment at that angle using the true linear equations or Bresenham's algorithm.

Furthermore, the skew operations dominate this approach to morphological operations
with rotated lines.

\par

Fortunately, there is a simpler, pixel accurate solution: we apply only a single
skew operation and correct for the change in length.

With this, line erosions at arbitrary angles between $[-\frac{\pi}{4},\frac{\pi}{4}]$ become (angles outside this range can be handled by flips and trasposes):

\begin{verbatim}
bits_erode_line(image,r,angle) {
    skew = tan(angle);
    corrected = r*cos(angle);
    bits_skew(image,skew);
    erode_line_horizontal(image,2*corrected);
    bits_skew(image,-skew);
}
\end{verbatim}

Here, the function \verb|bits_skew| moves each pixel 
at $(x,y)$ to $(x,y + \hbox{skew} \cdot x)$.

Note that the skew operation is perpendicular to the erosion.

It is likely faster to skew along each line (via bit shifting the line) 
and then perform the erosion perpendicular to that than the other way
around.

\section{Conversion Between Image Formats}

Conversion to/from run-length encoded representation to either
unpacked or packed bit-images is straight-forward.

We note that input/output can be implemented particularly efficiently
in terms of run-length image representations, since many binary image
formats internally already perform some form of run-length
compression, and their runs can be directly translated in runs in the
in-memory representation.

Even for file formats that do not use runs, input/output can be implemented
by compressing and decompressing the image one line at a time; that is,
we read a line by calling the image decompression library, decode the line
in an unpacked format into a 1D array, and then compute the run-length
encoding of that array.

\section{Connected Component Extraction and Statistics}

Connected components, and statistics over them, can also be computed
quickly:

\begin{itemize}

\item We associate a label value \verb|label[i][j]| with each run
  \verb|lines[i][j]|.

\item For each run in the entire image, we create a set in a
  union-find data structure.

\item We then iterate through all the lines in the image and, for each
  run in the current line merge its label with the labels of any runs
  in the line above.  This can be done in linear time in the number of runs
  in each line.

\item Finally, we renumber the entries in \verb|label[i][j]| according to
  the canonical set representative from the union-find data structure.

\end{itemize}

This is similar to a connected component algorithm on the line
ajacency graph (but the order in which nodes are explored can be different).

It is also similar to efficient connected component algorithms operating
on bitmap images, but runs are used instead of iterating over the pixels
or words.

\par

The output of this process is a set of runs \verb|lines[i][j]| and
corresponding labels \verb|label[i][j]|.

We can also think of these structures together as a run-length
compressed image with pixel values stored in the \verb|label| array.

For computing bounding boxes, centroids, moments, boundaries, boundary
properties, or other spatial statistics over these regions, we can
iterate through the runs and accumulate the corresponding information in
accumulator arrays indexed by the labels stored in the \verb|label| array.

\section{Other Operations}

There are a number of other operations that can be carried out quickly
on run-length representations:

\begin{itemize}
\item Run-length statistics are frequently used in document analysis to estimate
  character stroke widths, word spacings, and line spacings; they can be computed
  in linear time for both black and white runs by iterating through the runs
  of an image.  In the vertical direction, they can be computed by first
  transposing the image.
\item The line adjacency graph can be computed by treating the runs as
  nodes in the graph and creating edges between any runs in adjacent
  lines if the intervals represented by the runs overlap.
\item Standard skeletonization methods for the line adjaceny graph can
  be applied after computation of the LAG as described above.
  (See also \cite{piper92skeletons}.)
\item Run-length based extraction of lines and circles using the RAST
  algorithm \cite{keysers06rastrle} can be applied directly.
\end{itemize}

\section{Experiments}

We have implemented, among others, conversions between run-length,
packed bit, and unpacked bit representations of binary images,
transposition, all the morphological operations with rectangular
structuring elements described above, bitmap rotation by arbitrary
angles, computation of run-length statistics, connected component
labeling, and bounding box extraction.

For evaluating the general behavior of these algorithms and
determining whether they are feasible in practice, we are comparing
the performance of the run-length based algorithms with a companion
binary bit blit based morphology package, as well as the bitmap-based
binary morphology implementation in Leptonica 1.48 (8/30/07), an open
source morphological image processing library in use in production
code and containing well-documented algorithms and implementations
\cite{bloomberg02bits,leptonica-web}.

\par

Leptonica contains multiple implementations of binary morphology;
the fastest general-purpose implementation is {\tt pixErodeCompBrick} 
(and analogous names for other operations), a method that uses
separability and binary decomposition; it was used unless otherwise stated.

Leptonica also contains partially evaluated morphology operators for
a number of specific small mask sizes available under the names like
{\tt pixErodeBrickDwa}.  These were used when applicable.

Both libraries were compiled with their default (optimized) settings.

\par

In the experiments, we want to address several questions:

\begin{itemize}

\item What is the scaling behavior of the run length methods?

\item Which of the possible different run length implementations is better?

\item Is any one method uniformly better than the others, or do we need
    to perform algorithm selection?

\item How do these algorithms perform for the types of mask sizes and
    images found in typical document analysis tasks?

\end{itemize}

\subsection{2D Rectangular Masks}

\par

\begin{figure}[t]
\centerline{%
\includegraphics[width=3in]{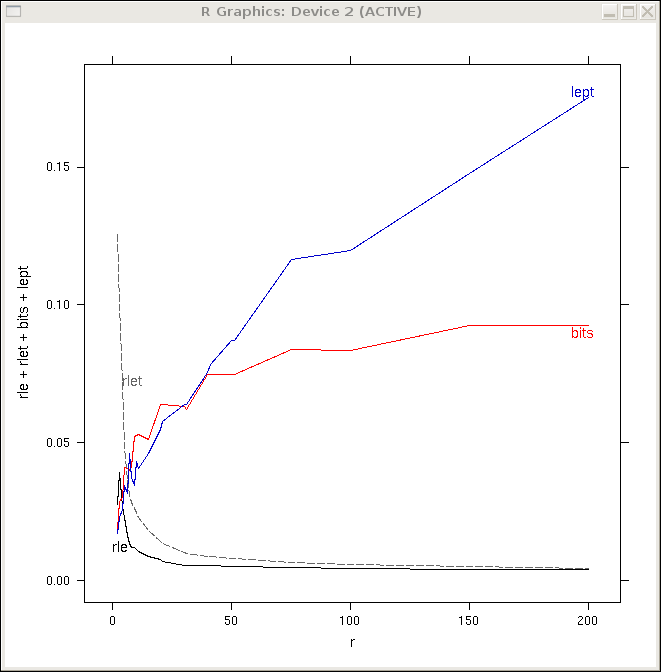}~%
\includegraphics[width=3in]{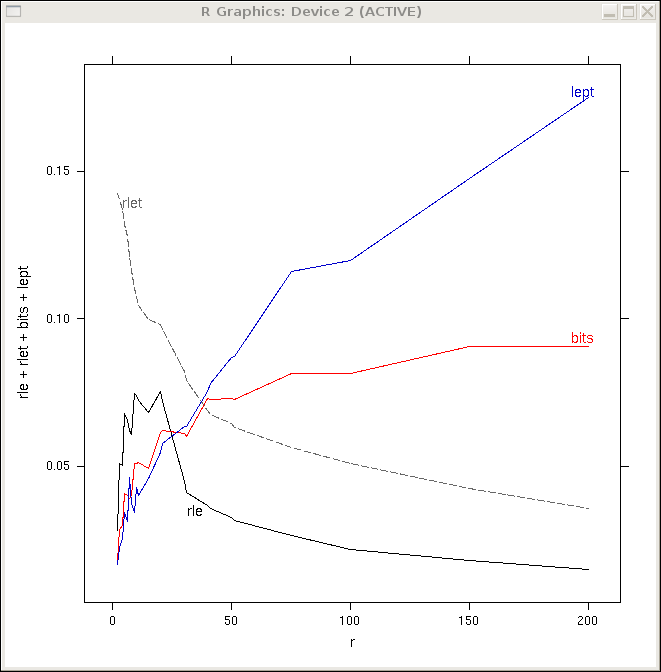}%
}
\caption{\label{fig-uw3}

Average running times (vertical axis) for opening (left) and closing (right) with square masks of
different size (horizontal axis) of the 1600 document image pages from the
UW3 document image database; the database consists of journal article
pages scanned at 300 dpi and binarized.  

{\it lept} = Leptonica bit blit based implementation (using {\tt
pixErodeBrickDwa} for sizes $<52$), {\it bits} = companion bit blit
library to the run length library, {\it rle} = run length-based morphology
using first within-line then between-line operations, {\it rlet} = run
length based morphology using within line operations only and transpose.

}
\end{figure}

\begin{figure}[t]
\centerline{%
\includegraphics[width=2.5in]{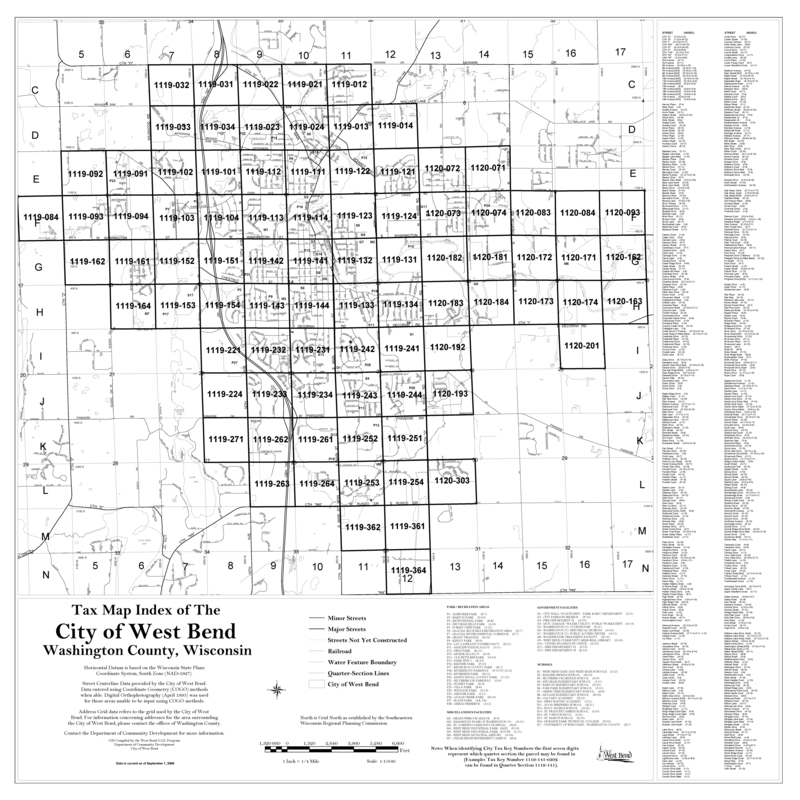}~%
}
\caption{\label{fig-cadastrali}
A $7000\times 7000$ image of a cadastral map used for performance
measurements.

}
\end{figure}

To gain some general insights into the behavior of the run length
methods for real-world document images, the running times of 
morphological operations on the 1600 images of the UW3 \cite{Phillips97}
database, 300 dpi binary images of scans of degraded journal publication
pages, were measured.

The results are shown in Figure~\ref{fig-uw3}.

We see that, except for masks of size five or below, the run length
implementation outperforms the bit blit implementation.

\par

By choosing at runtime between the bit blit implementation and the run length
implementation, we can obtain a method that shares the characteristics of both
kinds of images. 

As already noted above, the cross-over point can be determined automatically
either based on mask size and dpi, or based on output complexity.

This is shown as the bold curve in the figures; the curve does not
coincide the bit blit based running times because the run length figures
include the conversion times from run length representations to packed
bit representations and back to run length representations; in many
applications, these conversion costs can be eliminated.

By switching back to bit blit-based implementations for small mask sizes,
we can combine the two methods into a method that gives performance closer
to bit blit implementations at small sizes while still retaining the advantages
of run length methods at large sizes.

\par

\begin{figure}[t]
\centerline{%
\includegraphics[width=3in]{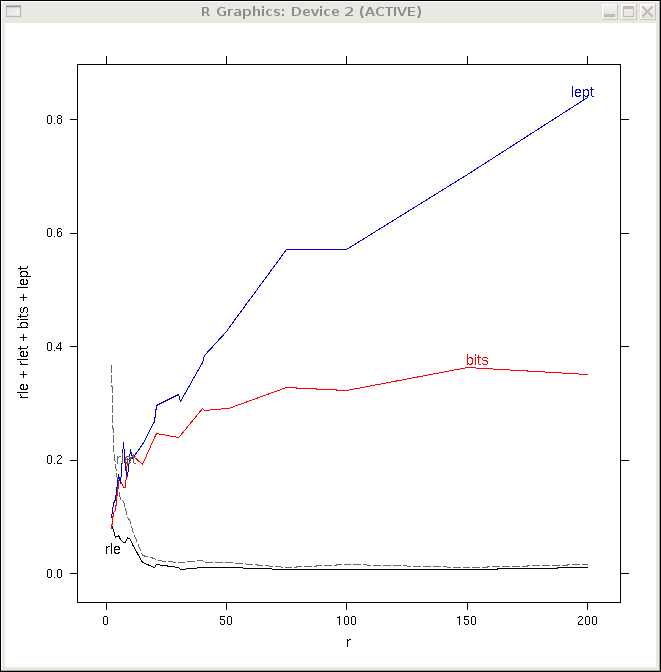}~%
\includegraphics[width=3in]{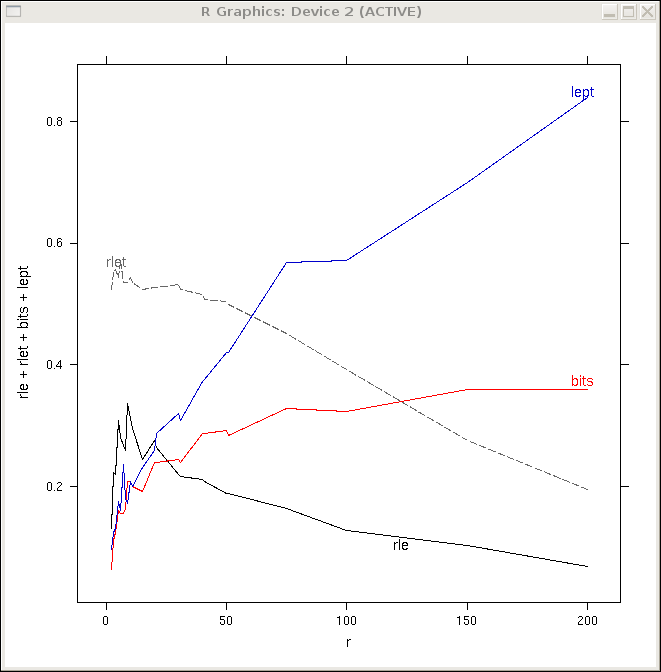}%
}
\caption{\label{fig-cadastral}

Times for opening (left) and closing (right) the $7000\times 7000$ image of a cadastral map in
Figure~\ref{fig-cadastrali} with the different sized square masked.

{\it lept} = Leptonica bit blit based implementation (using {\tt
pixErodeBrickDwa} for sizes $<52$), {\it bits} = companion bit blit
library to the run length library, {\it rle} = run length-based morphology
using first within-line then between-line operations, {\it rlet} = run
length based morphology using within line operations only and transpose.

}
\end{figure}

In a second experiment, we compared performance of the run length method
to Leptonica's bit-blit based morphology on a different document type
with a binarized $7000\times7000$ pixel cadastral map (Figure~\ref{fig-cadastral}).

\subsection{Circular Masks}

\begin{figure}[t]
\centerline{%
\includegraphics[width=3in]{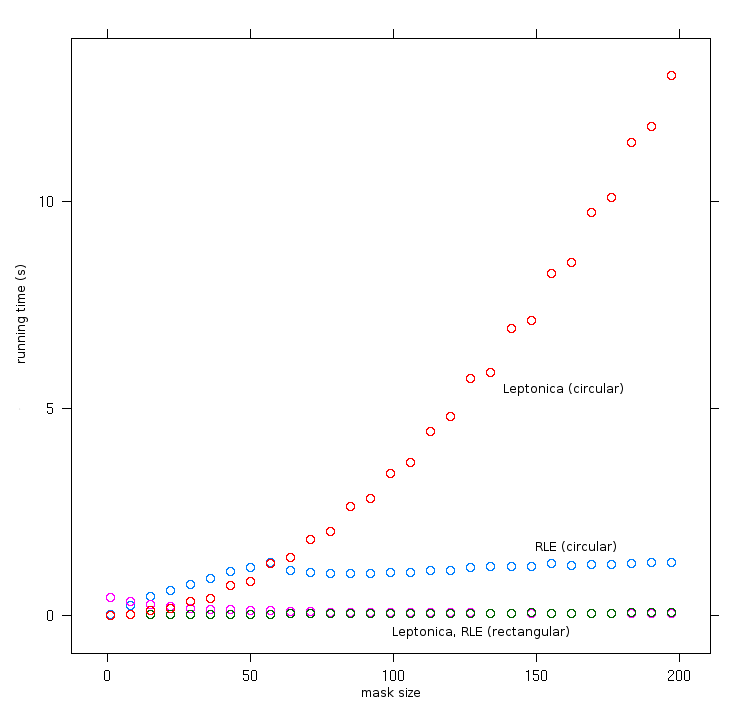}%
}
\caption{\label{fig-circle}

Times for morphological closing of a binary image of a two column text 
page at 600dpi using structuring elements of different size.  Both circular
structuring elements and square structuring elements are shown.

}
\end{figure}

As an illustration of the kind of performance achievable using the
run length methods for general purpose structuring elements,
Figure~\ref{fig-circle} shows the performance on circular and
rectangular structuring elements using run-length and brute force
bit blit morphology.

\par

As expected, the time for the brute force bit blit morphology grows
quadratically in the size of the structuring element.

Run length morphology grows linearly up to a point where the output
complexity (the number of runs in the output image) starts decreasing
and dominates the running time.

\subsection{Document Analysis Performance}

In the third experiment, we want to illustrate overall performance of
run length morphology methods as part of a simple morphological 
layout analysis system.

The method estimates the inter-word and inter-line spacing of document
images based on black and white run lengths, then performs erosion operations to
smear together connected components that are likely to be part of the
same blocks based on those estimates, and finally computes the bounding
boxes of the resulting large connected components; this approach is 
similar to the one in \cite{Wong82}

As the input, the 1600 pages from the UW3 database were
used.

These are 300dpi letter sized page images scanned from published journals.

\begin{figure}[t]
\centerline{%
\includegraphics[height=3in]{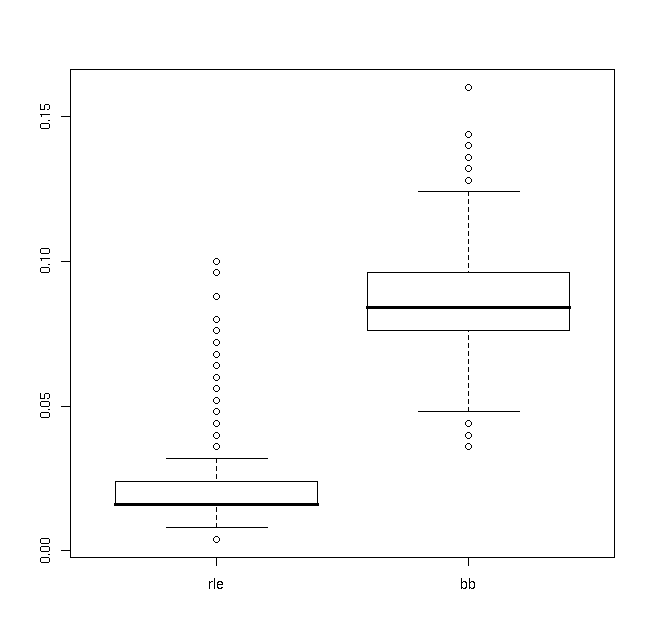}~%
\includegraphics[height=3in]{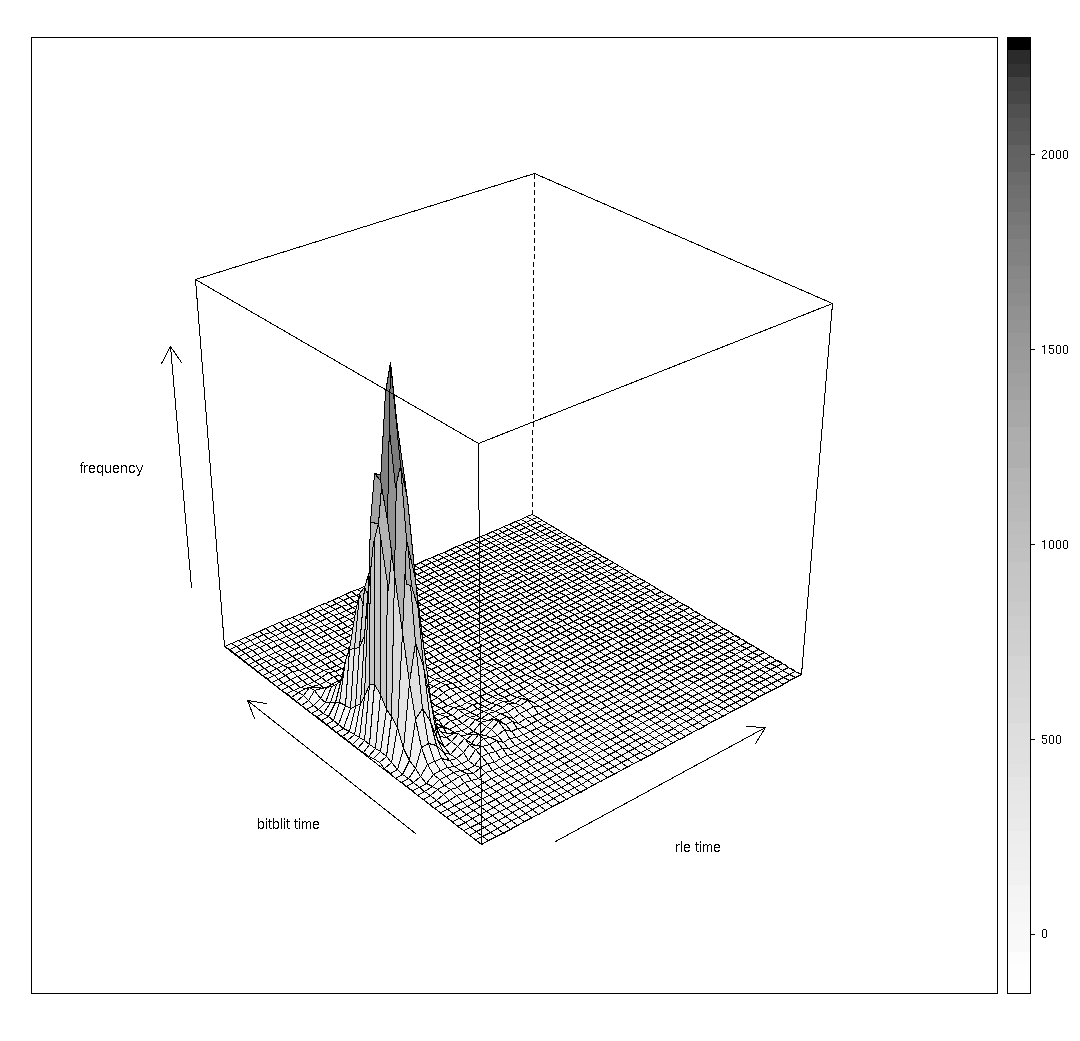}%
}
\caption{\label{fig-layouttime}

The left panel shows
boxplots of the running times of the morphological layout analysis system
using either the run length method (mixed within/between methods) 
or Leptonica's bit blit methods (using {\tt pixErodeBrickDwa} where possible).
The right panel shows a smoothed 2D histogram of the same data.
The system estimates character and line spacing from run lengths and then
performs a rectangular dilation that merges lines and characters into
blocks.  Finally, it computes bounding boxes of connected components.
Performance is shown over the 1600 IMGLINES images from the UW3
database.

}
\end{figure}

Relative performance of the run-length based method and Leptonica's
bit blit based method, including bounding box extraction, are shown
in Figure~\ref{fig-layouttime}.

The results show that run length morphological algorithms perform
about twice as fast at 300dpi than the bit blit based algorithms in
Leptonica (at 600dpi or 1200dpi, the advantage of run length methods
would be greater still).

\section{Discussion}

The paper describes a number of methods used in our open source binary
morphology library for performing morphological and related operations
on run length representations.

Although there is considerable overlap with previously published results,
some algorithmic details appear to be not well known, like the use of
skew operations for linear structuring elements, and the application of
run-length like decompositions and doubling for bit blit based morphology
with arbitrary structuring elements.

We hope that these methods and implementations will become a useful
reference implementation for other work, as well as a practical 
library for performing binary morphology on document images.

\par

The performance measurements on real-world document images over
a wide range of mask sizes, as well as the performance evaluation
in the context of a complete layout analysis system demonstrate
clearly that the run length morphology is an efficient alternative
to bit blit based morphology for realistic document images.

Furthermore, comparing 300dpi page images and performance on
larger cadastral maps, as well as theoretical considerations,
suggest that the advantage of run length methods increases as
the size and resolution of images increases.

\par 

The algorithms described in this paper were developed originally for
document image applications and have proven useful in a variety
of practical applications in the years since.

Although it is difficult to establish formally, generally, 
run length based algorithms seem to be somewhat easier to implement
efficiently than bit blits, since the boundary conditions and special
cases seem to be simpler and fewer.

Furthermore, run length morphology has no machine word size dependencies.

\par

It will remain for future work to see how the algorithms presented in this paper
relate to methods recently proposed in the literature.

Van Droogenbroeck \cite{mvd02rectangular}, for example, also describes
two algorithms using list or rectangular structures to aid in the computation
of fast binary morphology.

It appears that his methods are considerably more complex to
implement.  

Benchmarking and comparison of these new methods (including the ones
presented in this paper) will have to take into account both
space and running times, since both are crucial in practical
applications.

Unlike high performance bit blit methods, these new methods are
also sensitive to the complexity of both the input and output.

Van Droogenbroeck also raises the issue of implementation complexity;
while this is hard to quantify, it appears that the run length
methods described in this paper are easier to implement.

All these approaches are a trend to taking advantage of coherence in
images, similar to the way compression algorithms do.

\par

There are a number of obvious directions for extensions.

Operations on arbitrary morphological masks can be represented as run
length images as well, and morphology over them can be carried out using
similar approaches to those described here.

All the accesses within the inner loops of these algorithms are
sequential; this presents opportunities for more compact representations,
such as variable length multi-byte encodings of run lengths (since 
shorter run lengths tend to be more frequent than longer run lengths).
In fact, even Huffman coding or in-memory zlib compression for each
run are possible.

\par

Let us conclude by examining how these results can be used in practice.

Run length morphological operations can be incorporated into systems
in various ways: (1) a system can use RLE representations
and temporarily convert to bitmaps when there is a performance advantage
(taking into account the conversion costs), (2) a system can use bitmaps 
as its primary representation and temporarily switch to RLE when 
it speeds things up, (3) a system can keep everything in RLE format, or 
(4) a system can continue to keep everything in bitmap format.

It's clear that, provided the system selects the correct algorithm
automatically, (2) is no worse than (4) and that (1) no worse than (3)
in terms of performance, and the paper has, in fact, given examples
of speedups using such mixed approaches.

The experiments presented above on a simple layout analysis system also
suggest (but don't conclusively prove) that (3) may be faster than (4)
on average in real-world applications.  The question of whether (1) or
(2) is faster for real-world applications remains to be determined.

Many real-world imaging applications, such as printing engines, already
use run length representations internally, and the methods presented in
this paper give them the ability to integrate and perform morphological
operations directly and efficiently.

Existing bitmap-based libraries like Leptonica may want to choose
approach (2) to improve performance on large masks without affecting
software using the library.  Furthermore, the run length conversion
and operations can be incorporated directly into a blit-like operation,
resulting in a hybrid approach; this will be explored elsewhere.

Overall, run length approaches to binary morphology give us another useful
option for implementing morphological operations efficiently, in particular
in document imaging applications.

\section*{Acknowledgements}

The author would like Dan Bloomberg, Laurant Najman, Daniel Keysers,
and Christian Kofler for valuable suggestions and discussions.

\bibliographystyle{plain}
{\small \bibliography{rletr}}

\end{document}